\documentclass{PoS}

\usepackage{amsmath}
\usepackage{subfig}
\usepackage{epstopdf}
\usepackage{tikz}

\title{Upgrade of the T2K near detector ND280: effect on oscillation and cross-section analysis}

\ShortTitle{Upgrade of the T2K near detector ND280}

\author{\speaker{Mathieu Lamoureux}
        	~on behalf of the T2K collaboration \\
        SPP, IRFU, CEA Saclay,	Gif-sur-Yvette, France \\
        E-mail: \email{mathieu.lamoureux@cea.fr}}

\abstract{
The T2K neutrino oscillation experiment established the $\nu_\mu \rightarrow \nu_e$ appearance with only $10\%$ of the original beam request of $7.8\times10^{21}$ $30$\,GeV protons on target (POT). In view of the J-PARC program of upgrades of the beam intensity, the T2K-II proposal for $20\times10^{21}$\,POT aimed at establishing CP violation at $3\sigma$ level for a significant fraction of the possible $\delta_{CP}$ values. The Hyper-K proposal consists of a further increase by a factor $10$ of the far detector mass. Facing the potential increase of statistics by two orders of magnitude, it is of great importance to undertake a vigorous program of near detector upgrade, with the aim of reducing the overall statistical and systematic uncertainties at the appropriate level of better than 4\%. The design and performance of the proposed upgraded near-detector will be reported in these proceedings.
}

\FullConference{The 19th International Workshop on Neutrinos from Accelerators-NUFACT2017\\
		25-30 September, 2017\\
		Uppsala University, Uppsala, Sweden}

\begin{document}

\section{Introduction}

T2K \cite{T2K} is a long-baseline neutrino oscillation experiment in Japan. It uses a muon neutrino or antineutrino beam produced in Japan-Proton Accelerator Research Center (J-PARC) located in Tokai. 

There are three main detectors placed along the beamline:

\begin{itemize}
\item INGRID at $280$\,m from the target: it is composed of an iron-scintillator sandwich whose goal is to monitor the neutrino flux and its direction \cite{INGRID}.
\item ND280 at $280$\,m: it contains several sub-detectors all included in the UA1 magnet ($0.2$\,T). The main tracker volume consists of three Time Projection Chambers (TPCs) and two Fine-Grained Detectors (FGDs) in between. The FGDs are layers of scintillator bars which are used as an active target material, of around $1$\,ton each. There is also a $\pi^0$ detector (P0D) upsteam and the whole detector is surrounded by an electromagnetic calorimer (ECal).
ND280 is used to know the neutrino flux before any oscillation and constrain flux and cross-section model parameters.
\item Super-Kamiokande at $295$\,km: it is located $1$\,km deep in Kamioka mine. It is a 50 kton water Cherenkov detector ($22.5$\,kton fiducial volume). It can distinguish between muons and electrons and the energy reconstruction is made with the lepton kinematics assuming a charged current quasi-elastic interaction.
\end{itemize}

The latter two detectors are $2.5$\,degrees off-axis from the neutrino beam, allowing the experiment to have a narrow band muon neutrino beam peaked at $600$\,MeV, which corresponds to the first oscillation maximum at Super-Kamiokande.

The T2K collaboration has already measured muon (anti-)neutrino disappearance and electron (anti-)neutrino appearance \cite{nuAna}\cite{anuAna}. Recently, the experiment has also observed first hints of maximal CP violation \cite{CP}. 

\begin{figure}[!h]
\begin{minipage}[c]{0.45\linewidth}
\includegraphics[width=\linewidth]{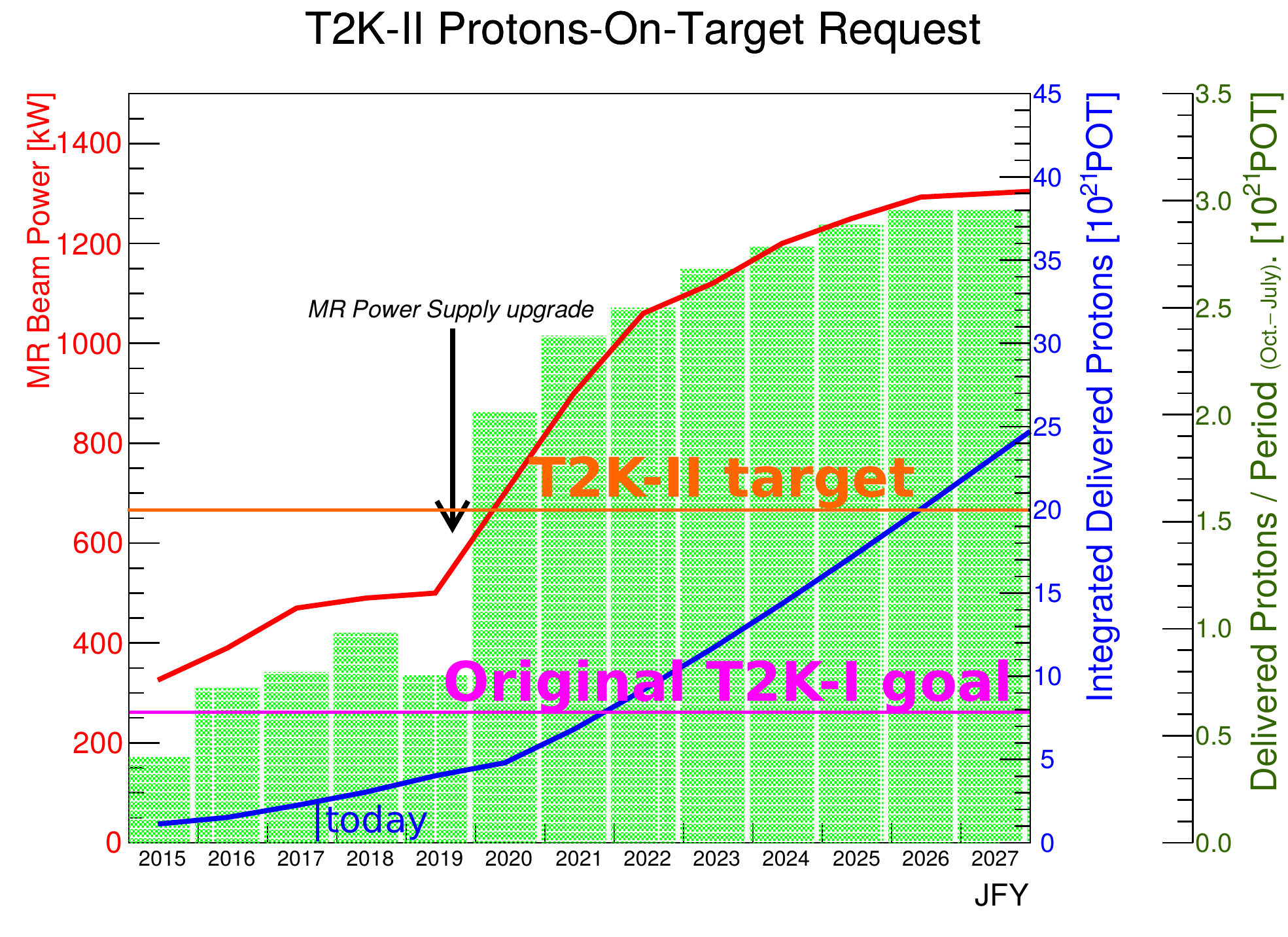} 

\caption{Expected beam power and Protons-on-Target accumulation. Plot taken from \cite{T2KII}. \label{T2K_II_POT}}
\end{minipage}
\hspace{.2in}
\begin{minipage}{0.55\linewidth}
\includegraphics[width=\linewidth]{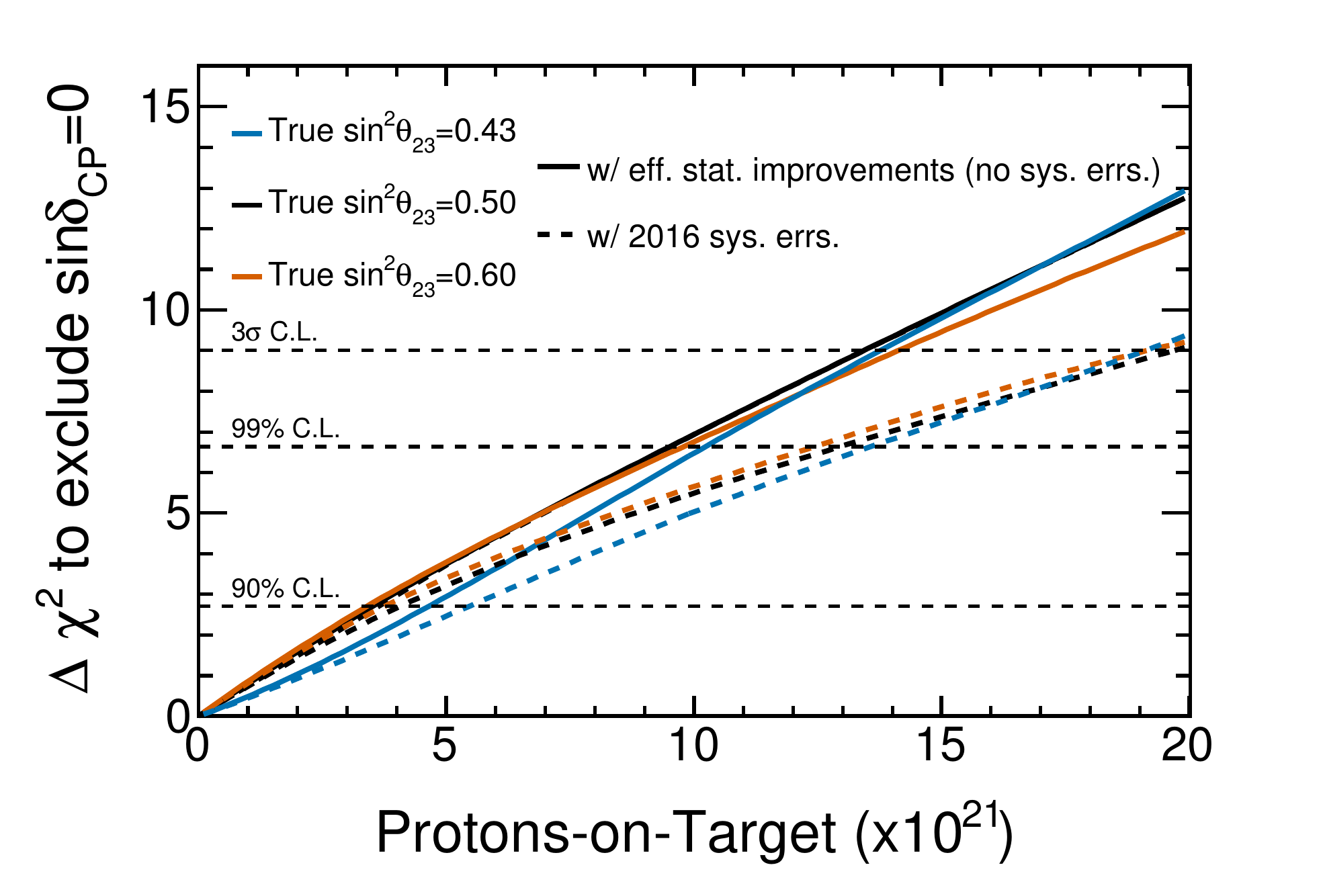}
\caption{Sensitivity to CP violation as a function of POT for T2K Phase 2 with predicted 2016 systematic errors. The $\delta_{CP}=-\frac{\pi}{2}$ and normal mass hierarchy are assumed to be true. Plot taken from \cite{T2KII}.\label{T2K_II}}
\end{minipage}
\end{figure}

T2K original goal was to reach a total of $7.8 \times 10^{21}$ protons on target (POT) in 2021, split between neutrino and antineutrino modes. An extension of the T2K running up to 2026, to achieve $20 \times 10^{21}$ protons on target, is under consideration \cite{T2KII}, as presented fig.~\ref{T2K_II_POT}. An upgrade of the J-PARC beam is scheduled, in order to reach a projected beam power of $1.3$\,MW.

The primary goal of this extension (T2K-II) is to achieve $>3\sigma$ sensitivity for CP violation in electron (anti-)neutrino appearance. Figure \ref{T2K_II} shows the expected significance of excluding CP conservation ($\delta_{CP} =0,\pi$) as a function of delivered POT for true $\delta_{CP} = -\pi/2$. It also clearly shows the impact of the current systematic errors on the sensitivity. 

An upgrade of ND280 is then needed to be able to reduce the flux and cross-section systematic uncertainties that can be constrained by the near-detector, in particular by increasing the angular acceptance of the particles created by neutrino interactions. This will help to better understand the latter.

\section{The proposed upgraded ND280 detector}

Current ND280 analyses have shown that the detector has only a good acceptance for forward tracks (with respect to neutrino direction), as shown in fig.~\ref{current}. The efficiency for high-angle (going perpendicular to the beam direction) tracks is low, due to the geometry of the FGD bars (vertical tracks only cross one or two bars, which is not sufficient for reconstruction) and to the fact there are no TPCs above and below the FGD. The identification of backward tracks is limited by track sense reconstruction that requires one to measure time-of-flight of particles.

In the proposed upgraded design \cite{upgrade}, as shown in fig.~\ref{upgrade}, a new tracker, consisting of a two-ton horizontal plastic scintillator target (about $1.8 \times 0.6 \times 2$\,m$^3$) sandwiched between two new horizontal TPCs, is built. This tracker would be surrounded by Time-of-Flight counters (made of plastic scintillator) to measure the direction of the tracks, reject the Out-of-Fiducial-Volume events and potentially be used for particle identification.

The total mass of active target for neutrino interactions increases from $2.2$\,tons (the two FGDs) to $4.3$\,tons (the two FGDs + the new target), allowing the doubling of the expected statistics for a given exposure.

\begin{figure}
\caption*{\emph{Colors:} \textcolor{orange}{current FGDs}, \textcolor{blue}{current TPCs}, \textcolor{gray}{P0D}, \textcolor{violet}{ECal}, \textcolor{magenta}{new target}, \textcolor{cyan}{new TPCs}, \textcolor{green!50!black}{new ToF counters}}

\begin{minipage}[c]{0.48\linewidth}
\centering
\begin{tikzpicture}[scale=0.72]
\draw[line width=1.5, -latex] (-6.5, 0)--(-5.8, 0) node[midway,above] {$\nu$};
\draw[-latex] (-6.8, -2)--(-6.5, -1.6) node[right] {x};
\draw[-latex] (-6.8, -2)--(-6.8, -1.4) node[above] {y};
\draw[-latex] (-6.8, -2)--(-6.2, -2) node[right] {z};

\fill[draw, fill=violet!50] (-5.7,-2.3) rectangle (-2.52,2.3);
\fill[draw, fill=violet!50] (-2.48,-2.3) rectangle (3.6,2.3);
\fill[draw, fill=black!30] (-5.7, -2) rectangle (-2.5, 2);
\fill[draw, fill=blue!30] (-2.5,-2) rectangle (-1, 2);
\fill[draw, fill=orange!30] (-1,-2) rectangle (-0.5, 2);
\fill[draw, fill=blue!30] (-0.5,-2) rectangle (1, 2);
\fill[draw, fill=orange!30] (1,-2) rectangle (1.5, 2);
\fill[draw, fill=blue!30] (1.5,-2) rectangle (3, 2);
\fill[draw, fill=violet!60] (3,-2) rectangle (3.6,2);
\fill[draw, white] (-3,-2.8) rectangle (3,-2.4);

\draw[line width=1.2, green!50!black, -latex] (-0.7,0.2)--(3.6, 0.5) node[midway,yshift=5pt,rotate=4,green!50!black] {\textbf{reconstructed}};
\draw[line width=1.2, red, -latex] (-0.7,0.2)--(-0.55, 2.3);
\draw[line width=1.2, red, -latex] (-0.7,0.2)--(-5.7, -1.8) node[midway,yshift=6pt,rotate=22] {\textbf{lower efficiency}};
\end{tikzpicture}

\caption{Current design of ND280 detector and its limitations.\label{current}}
\end{minipage}\hfill
\begin{minipage}[c]{0.48\linewidth}
\centering
\begin{tikzpicture}[scale=0.72]
\draw[line width=1.5, -latex] (-6.5, 0)--(-5.8, 0) node[midway,above] {$\nu$};
\fill[draw, fill=violet!50] (-5.7,-2.3) rectangle (-2.52,2.3);
\fill[draw, fill=violet!50] (-2.48,-2.3) rectangle (3.6,2.3);
\fill[draw, fill=black!30] (-5.7, -2) rectangle (-5.5, 2);
\fill[draw, fill=magenta!30] (-5.5, -0.5) rectangle (-2.5, 0.5);
\fill[draw, fill=cyan!30] (-5.5, -0.5) rectangle (-2.5,-2);
\fill[draw, fill=cyan!30] (-5.5,  0.5) rectangle (-2.5, 2);
\fill[draw, fill=blue!30] (-2.5,-2) rectangle (-1, 2);
\fill[draw, fill=orange!30] (-1,-2) rectangle (-0.5, 2);
\fill[draw, fill=blue!30] (-0.5,-2) rectangle (1, 2);
\fill[draw, fill=orange!30] (1,-2) rectangle (1.5, 2);
\fill[draw, fill=blue!30] (1.5,-2) rectangle (3, 2);
\fill[draw, fill=violet!60] (3,-2) rectangle (3.6,2);
\draw[line width=1.5, green!50!black] (-5.5, -2) rectangle (-2.5, 2);
\fill[draw, white] (-3,-2.8) rectangle (3,-2.4);

\draw[line width=1.2, green!50!black, -latex] (-0.7,0.2)--(3.6, 0.5);
\draw[line width=1.2, green!50!black, -latex] (-4,0.2)--(-3.9, 2.3);
\draw[line width=1.2, green!50!black, -latex] (-0.7,0.2)--(-5.7, -1.8);
\end{tikzpicture}

\caption{Proposed upgraded design of ND280 detector.\label{upgrade}}
\end{minipage}
\end{figure}
 
\clearpage 
 
\begin{figure}
\centering
\includegraphics[width=0.48\linewidth]{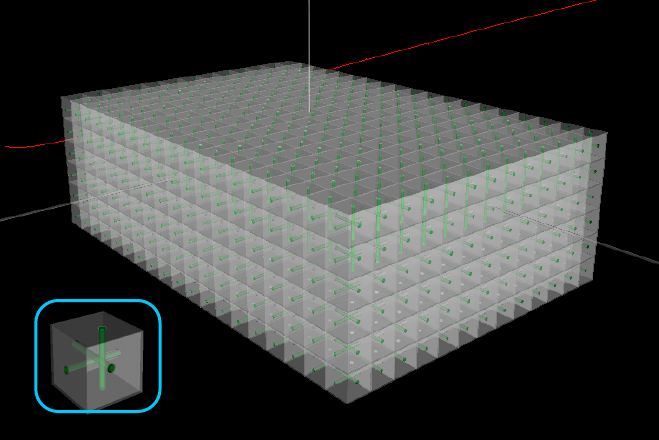} \hspace{0.3cm}
\includegraphics[width=0.46\linewidth]{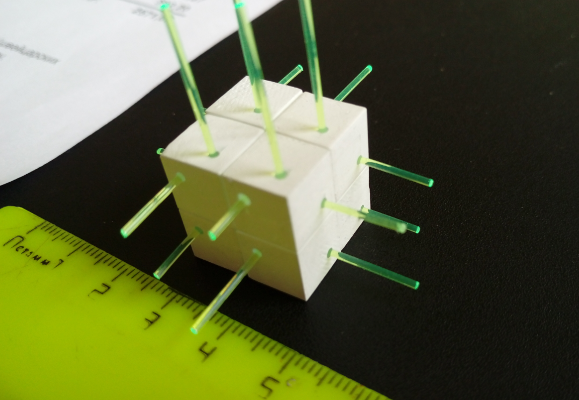} 
\caption{Left: Schematic view of Super-FGD detector. Right: picture of a small prototype of the Super-FGD, taken from \cite{SuperFGD}.}
\label{SuperFGD}
\end{figure}

Different technologies for the new scintillating target are under consideration:

\begin{itemize}
\item The first option is a new type of detector called Super-FGD \cite{SuperFGD}, consisting of a matrix of $1 \times 1 \times 1$\,cm$^3$ cubes made of extruded plastic scintillator. Each cube is crossed by three wavelength shifting (WLS) fibers along the three directions X, Y and Z, as shown in fig.~\ref{SuperFGD}. R\&D is ongoing inside the collaboration. 

In this configuration, each hit results in light in the three fibers (giving at least three times more light-yield than a classic scintillator). This provides three views (XZ, YZ, XY) to be used to reconstruct neutrino events, even with low-momentum tracks in any direction (for instance, the proton threshold is expected to reduce from $450$\,MeV/c in the current FGD to $300$\,MeV/c in the Super-FGD). This detector could also provide a better separation of electrons and converted photons than the current FGD.

\item The second option consists of relying on the current FGD known technology, by building a horizontal FGD (scintillator bars in X and Z directions), in order to detect particles propagating perpendicularly to the beam, even though such a detector provides only two views and has limited forward acceptance.
\end{itemize}

\section{Performance}

\begin{figure}
\centering
\includegraphics[width=0.75\linewidth]{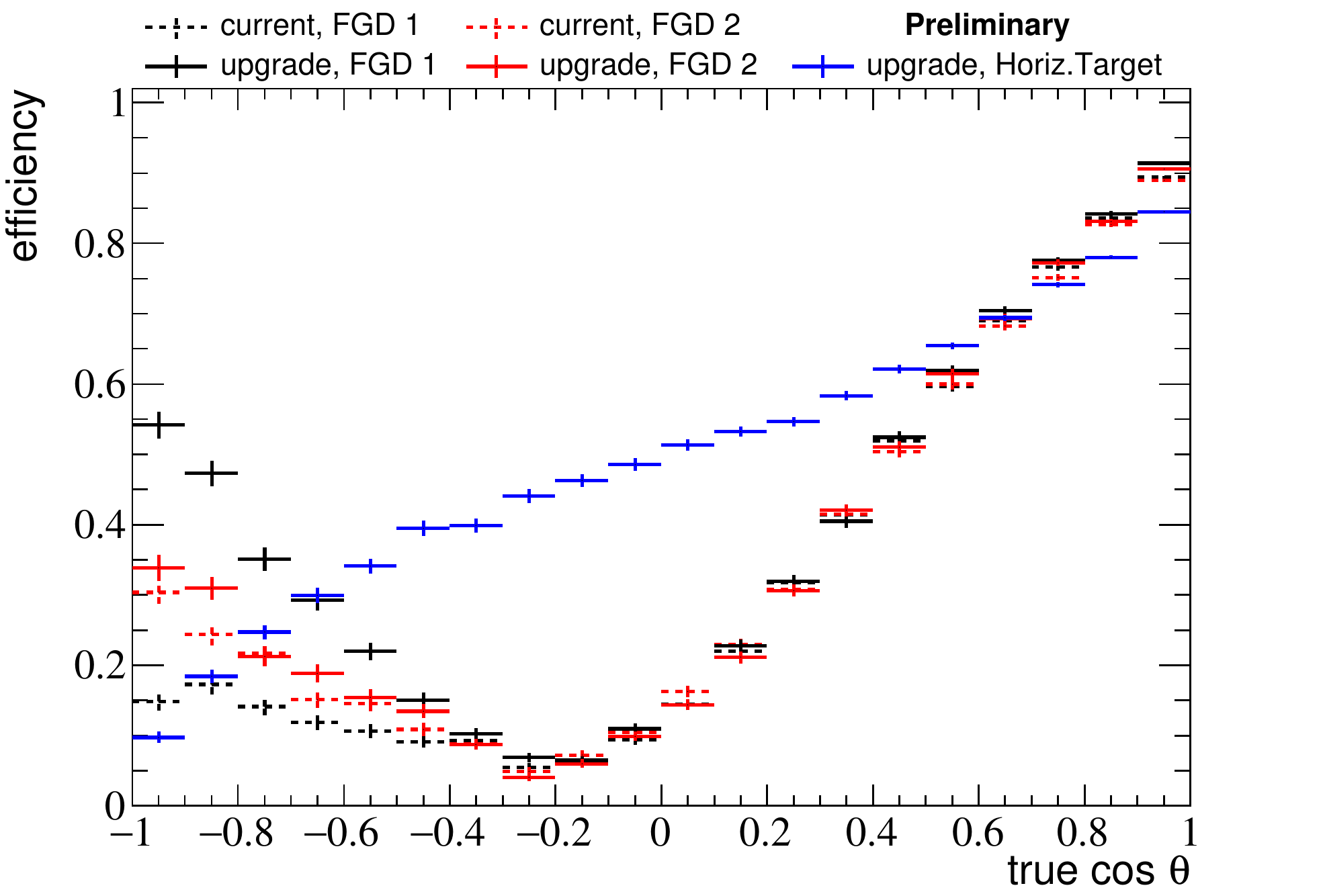} 
\caption{Selection efficiency of $\nu_{\mu}$CC events as a function of the muon polar angle, with muons detected in TPC. The solid lines correspond to ND280 upgrade configuration and the dashed lines correspond to current-like configuration. Plot taken from \cite{upgrade}.}
\label{eff}
\end{figure}

Simulations with GEANT4 \cite{GEANT4} and GENIE \cite{GENIE} have been performed in order to compare the performance of the current-like detector and the proposed upgraded configuration. Figure~\ref{eff} shows the selection efficiency of $\nu_{\mu}$ charged-current (CC) interactions, as a function of muon angle. We see a large improvement for muons emitted at high-angle (perpendicular to the beam direction) in the new target, thanks to the two horizontal TPCs, and backward in the closest FGD, thanks to timing information provided by Time-of-Flight counters.

This study only considers muons reconstructed and identified in a TPC. Performance may improve further by considering muons stopping in the target (as the SuperFGD is expected to provide an acceptance in all directions that is higher than $90\%$).

Based on these efficiencies, sensitivity studies are done in order to assess the impact of the upgrade on oscillation and physics analyses. Table \ref{BANFF} presents the improvement on some of the systematic uncertainties of the neutrino oscillation measurement.

\begin{table}[!tbp]
	\centering
	\caption{Sensitivity to some flux and cross-section parameters of interest for the current ND280 and the upgrade configuration. Values taken from \cite{upgrade}.}
	\begin{tabular}{c|c|c}
		\hline
		\hline
		Parameter 					& Current-like ($\%$) 	& Upgrade-like ($\%$) \\
		\hline
		SK flux normalisation  			& $2.9$ 			& $2.1$ \\ 
		($ 0.6 < \text{E}  < 0.7$\,GeV) 
		MA$_{QE}$ (GeV/c$^2$) 			& $2.6$ 			& $1.8$ \\ 
		$\nu_{\mu}$ 2p2h normalisation	& $9.5$				& $5.9$ \\ 
		2p2h shape on Carbon 			& $15.6$ 			& $9.4$ \\ 
		\hline
		\hline
	\end{tabular}
	\label{BANFF}
\end{table}

\section{Conclusion}

T2K plans to extend its data taking up to 2026, followed by the Hyper-K project \cite{HyperK} where the mass of the far detector will be increased by a factor $10$. In this context, a near-detector upgrade is a necessary step in order to reduce the uncertainty on the neutrino event rate prediction at the far detector and look for the first evidence of CP violation in the leptonic sector.

An upgraded design, consisting of a new plastic scintillator detector surrounded by two horizontal TPCs, is proposed. An Expression Of Interest was submitted to CERN SPSC in January 2017 (CERN-SPSC-2017-002 and SPSC-EOI-015). The R\&D is ongoing, for both the TPCs and SuperFGD, a novel scintillator detector.

First performance studies have shown that it is possible to cover better the phase space of neutrino interactions and this allows one to reduce the uncertainties on parameters used in oscillation analysis. Studies of $\nu_e$ interactions and electron-photon separation in new target are ongoing.

All these improvements, with respect to the current detector design, will help to better understand and constrain neutrino interaction models.

\end{document}